\documentclass[Physsubmission, Phys]{SciPost}

\hypersetup{
    colorlinks,
    linkcolor={red!50!black},
    citecolor={blue!50!black},
    urlcolor={blue!80!black}
}

\usepackage[bitstream-charter]{mathdesign}
\usepackage{amsmath}
\usepackage{bm}

\DeclareSymbolFont{usualmathcal}{OMS}{cmsy}{m}{n}
\DeclareSymbolFontAlphabet{\mathcal}{usualmathcal}

\usepackage{cite,./mcite}

\begin{document}

\begin{center}{\Large \textbf{
Drell-Yan transverse spectra at the LHC: a comparison of parton branching and analytical resummation approaches
}}\end{center}

% TODO: write the author list here. Use initials + surname format.
% Separate subsequent authors by a comma, omit comma at the end of the list.
% Mark the corresponding author with a superscript *.
\begin{center}
Aron Mees van Kampen\textsuperscript{1},
\end{center}

\begin{center}
{\bf 1} Elementaire deeltjesfysica, University of Antwerp
\\
* AronMees.vanKampen@uantwerpen.be
\end{center}

\begin{center}
\today
\end{center}

% For convenience during refereeing (optional),
% you can turn on line numbers by uncommenting the next line:
%\linenumbers

\definecolor{palegray}{gray}{0.95}
\begin{center}
\colorbox{palegray}{
  \begin{tabular}{rr}
  \begin{minipage}{0.1\textwidth}
    \includegraphics[width=22mm]{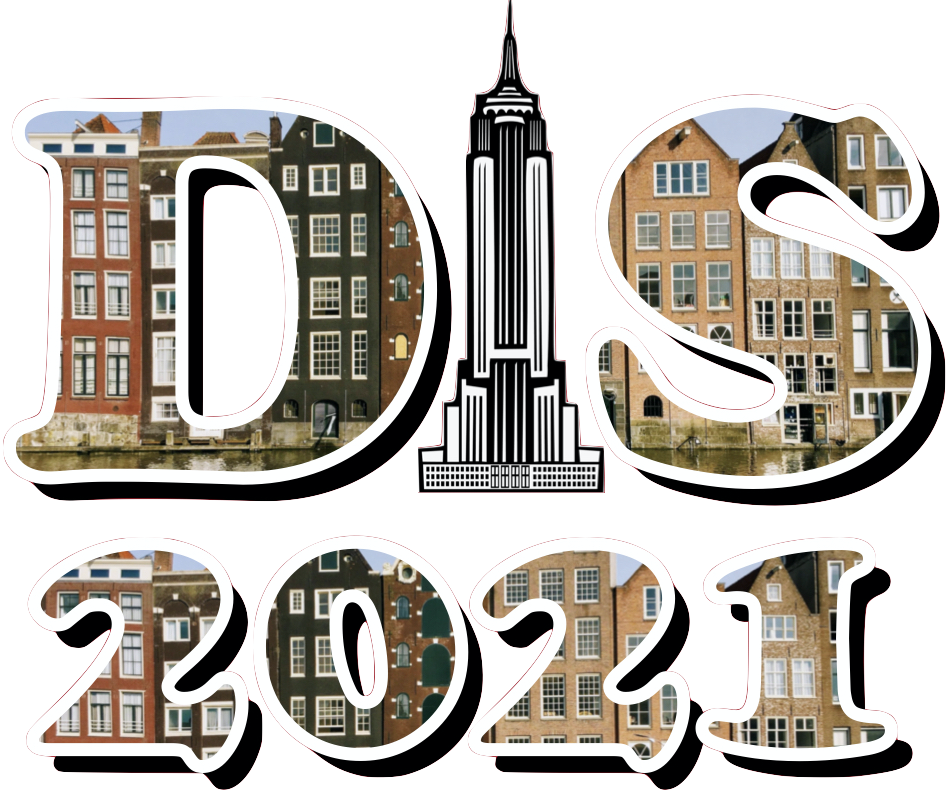}
  \end{minipage}
  &
  \begin{minipage}{0.75\textwidth}
    \begin{center}
    {\it Proceedings for the XXVIII International Workshop\\ on Deep-Inelastic Scattering and
Related Subjects,}\\
    {\it Stony Brook University, New York, USA, 12-16 April 2021} \\
    \doi{10.21468/SciPostPhysProc.?}\\
    \end{center}
  \end{minipage}
\end{tabular}
}
\end{center}

\section*{Abstract}
{\bf
% TODO: write your abstract here.
I report on precision comparisons of  theoretical predictions  
for Drell-Yan (DY) transverse momentum spectra from two approaches: the analytical resummation approach based on 
Collins-Soper-Sterman (CSS) formalism and the parton branching (PB) approach to the evolution of transverse momentum 
dependent  (TMD) parton distributions. 
%To increase precision of LHC observable predictions, QCD radiation causing large logarithms at all orders of $\alpha_s$ and non-perturbative (NP) effects need to be described. For this, a multitude of formalisms implementing transverse momentum dependent (TMD) factorization has been developed. We analytically and numerically compare the resummation approach by CSS with the parton branching (PB) method for Monte Carlo event generation with TMDs. We observe that PB reaches NLL accuracy in the Sudakov, predictions of the Drell-Yan $p_T$ spectrum show good agreement at low-$p_T$ and interesting features arise in comparing scale variations and NP parameterization.
}

%\vspace{10pt}
%\noindent\rule{\textwidth}{1pt}
%\tableofcontents\thispagestyle{fancy}
%\noindent\rule{\textwidth}{1pt}
%\vspace{10pt}

\section{Introduction}
\label{sec:intro}
Extensions of the QCD collinear factorization theorem~\cite{Collins:1989gx} involving 
TMD parton distributions~\cite{Angeles-Martinez:2015sea} are relevant to perform perturbative resummations to all 
orders in the strong coupling in kinematic regions of interest to  collider physics, 
e.g.~the Sudakov (or low-$p_T$) region~\cite{Collins:1981va,Collins:1984kg,Collins:2011zzd	} and 
high-energy (or low-$x$) region~\cite{Catani:1990xk,Catani:1990eg,Catani:1994sq}. 

In the last few years, precision measurements of transverse momentum spectra in DY lepton-pair production have been 
carried out at the Large Hadron Collider (LHC)~\cite{Aad:2015auj,CMS:2016mwa,ATLAS:2019zci,CMS:2019raw}, which call 
for detailed studies of TMD resummations in the region of low vector-boson transverse momentum $p_T$. 

In this article I report on a comparison of theoretical predictions for DY $p_T$ spectra based on two different 
approaches: the  analytical resummation  approach based on the CSS formalism~\cite{Collins:1981va,Collins:1984kg} and the 
PB approach~\cite{Hautmann:2017fcj,Hautmann:2017xtx} to the evolution of TMD distributions. I start in Sec.~\ref{sec:theory} by briefly recalling the basic elements of the two methods. In Sec.~\ref{sec:analytical} I compare them at analytical level. In Sec.~\ref{sec:numerical} I present numerical comparisons. I give conclusions in Sec.~\ref{sec:conclusion}. 

\section{Theoretical methods}
\label{sec:theory}
\subsection{CSS method}
The CSS method~\cite{Collins:1981va,Collins:1984kg} decomposes the DY differential cross section into the sum of the resummed term (dominant for $p_T \ll Q$, where $Q$ is the vector boson invariant mass) and the finite term (important for $p_T$ of order $Q$). The resummed term is factorized in impact parameter space into a hard-scattering coefficient function  and TMD distributions, which in turn are expressed as convolutions of Sudakov form factors, collinear evolution coefficients and parton distributions. 

The resummed and finite terms require an appropriate \textit{matching} procedure in order to avoid double counting. The matching can be done by subtracting counterterms from the resummed cross section~\cite{Catani:2000vq,deFlorian:2000pr,Bozzi:2005wk}. 
 
Nonperturbative effects are incorporated in the formalism through a Gaussian smearing factor in impact parameter space, as well as nonperturbative contributions to Sudakov evolution~\cite{Ladinsky:1993zn,Landry:2002ix,Konychev:2005iy,Hautmann:2020cyp}. 

\subsection{PB method}
The parton branching (PB) method~\cite{Hautmann:2017fcj,Hautmann:2017xtx} 
formulates the evolution of TMD distributions in terms of Sudakov form factors and real emission 
splitting functions using the \textit{unitarity} picture of parton interactions commonly employed in 
showering Monte Carlo algorithms~\cite{Bengtsson:1987kr,Marchesini:1987cf}. The resolvable and non-resolvable 
regions of the branching phase space are separated by the soft-gluon resolution scale $z_M$ 
(in general, branching scale dependent~\cite{Hautmann:2019biw}). Soft-gluon angular ordering is implemented in the TMD evolution so that color coherence effects are taken into account and the collinear limit agrees with the coherent branching approach of Refs.~\cite{Marchesini:1987cf,Catani:1990rr}.

The matching with finite-order hard-scattering   matrix elements is not done additively as in the CSS framework, but rather by using the showering operator method with subtracted matrix elements~\cite{Collins:2000gd,BermudezMartinez:2020tys,Martinez:2019mwt}. 

Nonperturbative TMD effects are encaptured in the TMD distribution at the starting evolution scale $\mu_0 = {\cal O} $(1 GeV), which is parametrized (for instance, as a transverse momentum Gaussian times a  longitudinal momentum distribution in a simplest parametrization) and determined from fits to experimental data~\cite{Martinez:2018jxt}. 

\section{Analytical comparison}
\label{sec:analytical}
The Sudakov form factor in CSS is written as 
\begin{equation} \label{eq:CSS_Sudakov}
S(Q,b) = \exp \left( - \int_{b_0/b^2}^{Q^2} \frac{d\mu^2}{\mu^2} \left[ \ln \left( \frac{Q^2}{\mu^2}\right) A(\alpha_s(\mu^2)) + B(\alpha_s(\mu^2)) \right] \right) 
%\times S_{NP}(b),
\end{equation}
where $b$ is the impact parameter, and the functions $A$ and $B$ are perturbatively calculable as power series expansions in $\alpha_s$, 
$ A = \sum_n A^{(n)} \alpha_s^n$,  $ B = \sum_n B^{(n)} \alpha_s^n$, and control  
the resummation of double-logarithmic and single-logarithmic perturbative corrections respectively, with  
$A^{(1)}$ corresponding to leading logarithms (LL),  $A^{(2)}$ and $B^{(1)}$ corresponding to next-to-leading logarithms (NLL), and so on. 

The Sudakov form factor in PB is written as 
\begin{equation} \label{eq:PBSudakov}
\Delta_a(\mu^2,\mu_0^2) = \exp \left( -\sum_b \int_{\mu_0^2}^{\mu^2}\frac{d\mu'^2}{\mu'^2}\int_0^{z_M}dz \ z \ P_{ba}^{(R)}(\alpha_s(q_t),z) \right) \; , 
\end{equation}
where $z_M$ is the soft-gluon resolution scale and $ P_{ba}^{(R)}$ are the real emission splitting functions,  
perturbatively calculable as power series expansions in $\alpha_s$. By rewriting  
the PB Sudakov  using the momentum sum rule, virtual splitting functions and the angular ordering condition,  we can compare 
the CSS perturbative coefficients  $A^{(i)}$  with the PB coefficients $k^{(i-1)}$ of Ref.~\cite{Hautmann:2017fcj}, and 
the CSS perturbative coefficients  $B^{(i)}$  with the PB coefficients $d^{(i-1)}$ of Ref.~\cite{Hautmann:2017fcj}. 

At LL and NLL level, all coefficients are in agreement between CSS and PB 
\cite{AMvanKampen-et-al_work-in-preparation}. At next-to-next-to-leading logarithms (NNLL), differences 
between coefficients both at the single and double logarithmic level are observed which can be understood~\cite{AMvanKampen-et-al_work-in-preparation} 
either  as a result of renormalization group transformations and 
resummation scheme dependence \cite{Catani:2000vq,deFlorian:2000pr} (in the single-log coefficients) or as an effect of 
soft-gluon effective coupling \cite{Catani:2019rvy,Banfi:2018mcq,Becher:2010tm} (in the double-log coefficients).  
This gives the possibility of future systematic comparisons of precision predictions in the phase space regions influenced by Sudakov resummation. 

\section{Numerical comparison} \label{sec:numerical}
First numerical comparisons have been performed for the DY transverse momentum spectrum based on the implementation of the CSS method in the \textsc{reSolve} \cite{Coradeschi:2017zzw} event generator and on the implementation of the PB method in the \textsc{Cascade}  \cite{Baranov:2021uol,Jung:2010si} event generator, supplemented by the uPDFevolv program \cite{Hautmann:2014uua} for TMD evolution. 

\begin{figure}[h]
\centering
\includegraphics[width=0.45\textwidth]{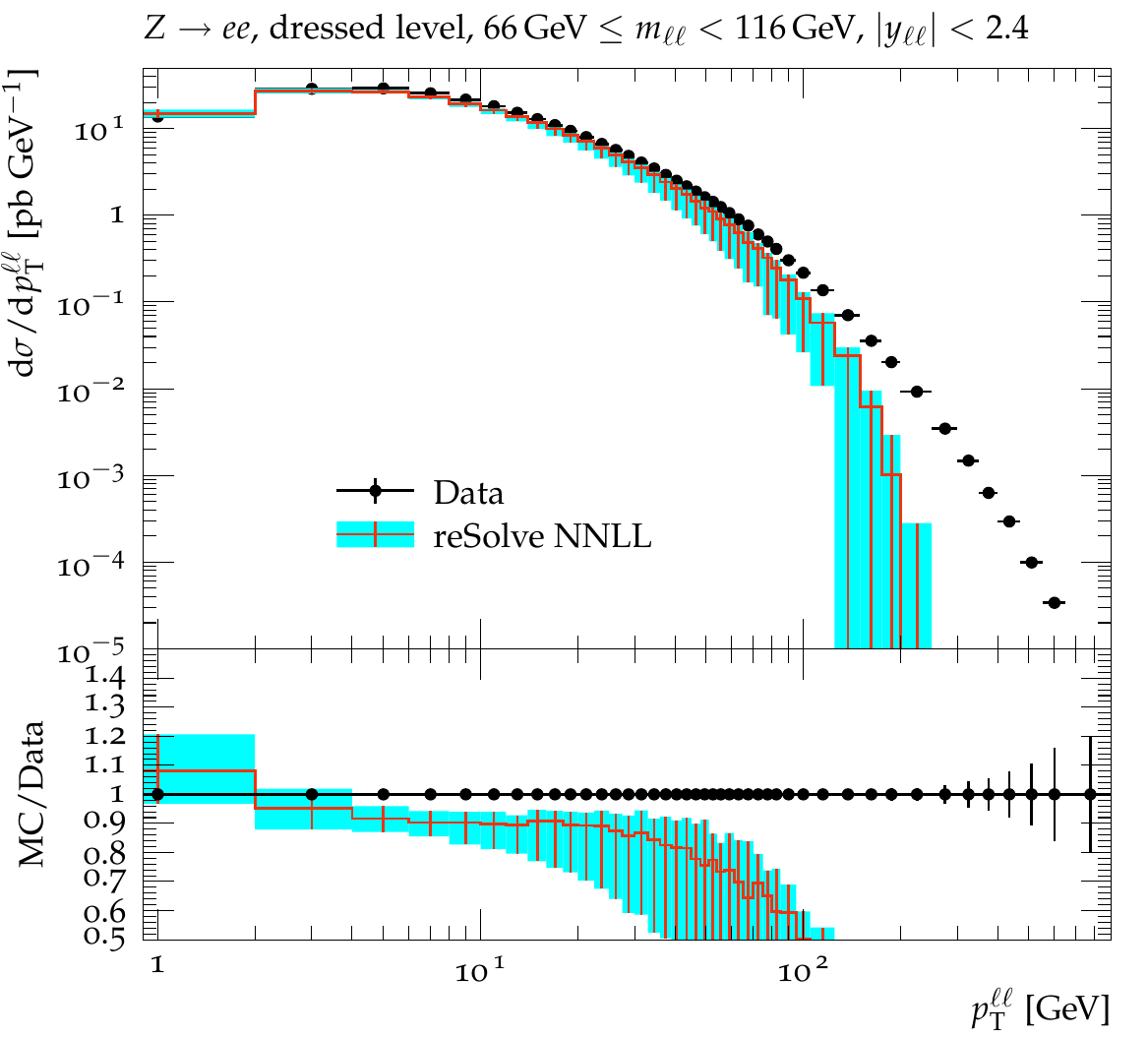}  \hspace{0.2cm} \includegraphics[width=0.45\textwidth]{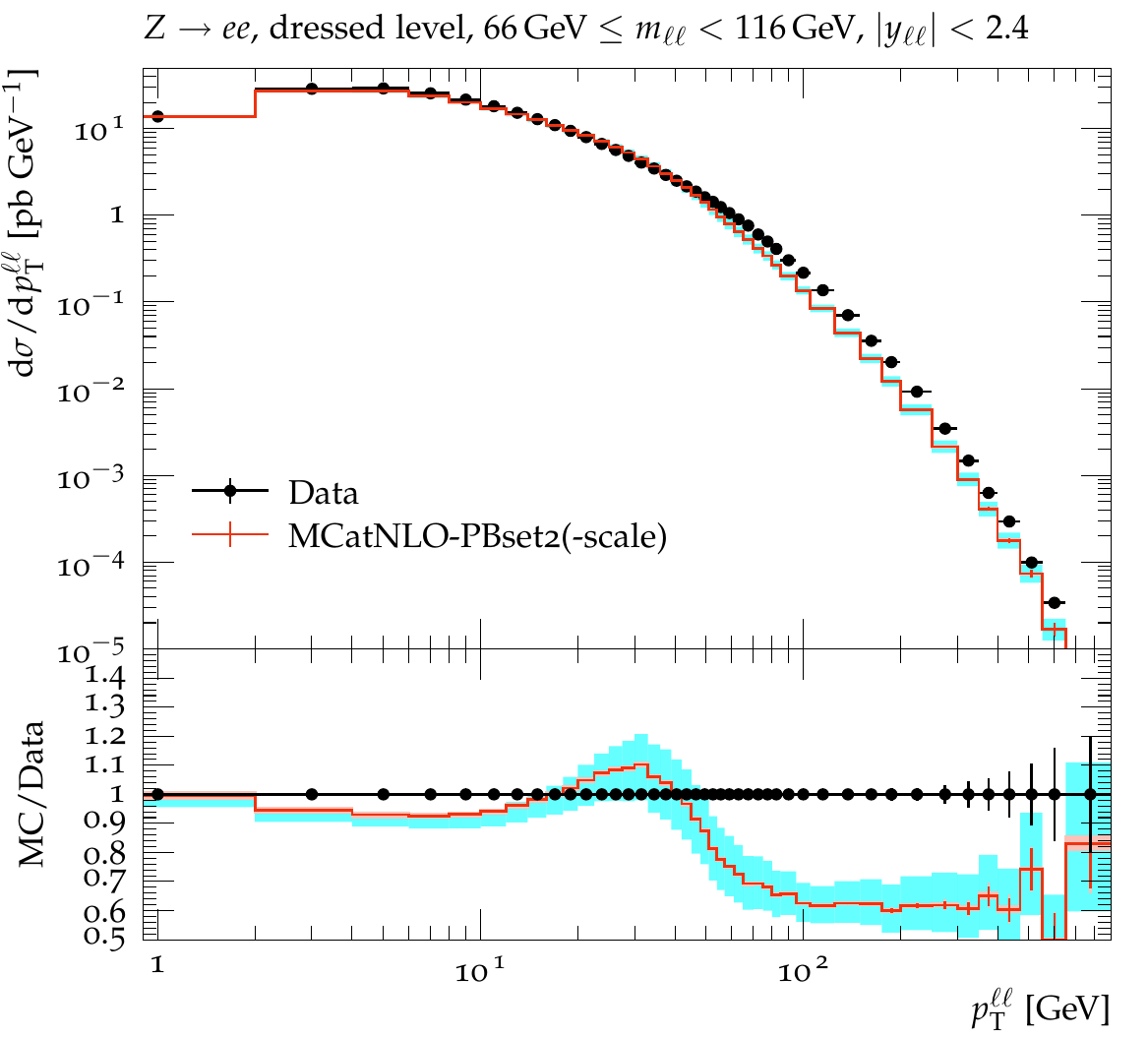} 
\caption{Inclusive Z boson $p_T$ spectrum by reSolve/CSS (left) and Cascade/PB (right). Theoretical uncertainty bands are shown in blue.}
\label{fig:ZpTnumerical}
\end{figure}

\textsc{reSolve} implements the resummed part of the cross section as described by CSS up to 
NNLL accuracy  \cite{Coradeschi:2017zzw,Accomando:2019ahs}. Matching of the resummed part to finite-order terms is not included. The result for the $p_T$ spectrum is shown in the left plot of Fig. \ref{fig:ZpTnumerical}. 

The calculation with  \textsc{Cascade} implements the PB TMD, and performs the matching with
NLO hard-scattering events from \textsc{Madgraph\_aMC@NLO} \cite{Alwall:2014hca}, in combination with the TMD. For this calculation the PB-NLO-HERAI+II-2018-set2 TMD PDF \cite{Martinez:2018jxt} is extracted from \textsc{TMDlib} \cite{Abdulov:2021ivr,Hautmann:2014kza} and used by \textsc{Cascade}. The matching of the NLO events to the TMD evolution is done as in Ref. \cite{Martinez:2019mwt} by using \textsc{Herwig6} subtraction terms \cite{Corcella:2002jc} and a matching scale to avoid double counting. 
The result with PB is shown in the right plot of Fig. \ref{fig:ZpTnumerical}.

From the plots, a few observations can be made. 
Firstly, both approaches have an accurate prediction in the low-$p_T$ region up to $p_T \sim 20$ GeV. The predictions fail to describe the high-$p_T$ region for different reasons. In \textsc{reSolve} it is due to absence of finite-order terms. The PB TMD is matched to NLO hard-scattering events for inclusive Z production. The deficit at high-$p_T$ is due to missing higher orders or higher jet multiplicities. One way to increase accuracy for the high-$p_T$ tail is to include TMD \textit{merging} with higher jet multiplicities, as in Ref.~\cite{Martinez:2021chk}. 

The second observation is that the theory uncertainties are calculated differently. In \textsc{reSolve} they are obtained by varying three scales: the renormalization ($\mu_R$), factorization ($\mu_F$) and resummation ($\mu_S$) scales. Since the resummation scale is mostly important for matching to finite terms, the variation of $\mu_S$ gives a distorted view of the uncertainty at the intermediate $p_T$ region. The PB uncertainties in Fig. \ref{fig:ZpTnumerical} include only two scale variations, that of $\mu_R$ and $\mu_F$. The matching scale has not been varied yet. Besides these uncertainties, there are also TMD uncertainties. In this kinematic range of energy, mass and transverse momentum, these are much smaller than the scale variations.

Finally, the nonperturbative contributions are different. In the CSS implementation of reSolve, the nonperturbative factor is an overall Gaussian smearing factor of the form $\exp(-gb^2)$ \cite{Coradeschi:2017zzw}. In PB, the starting TMD contains all the nonperturbative parts. The intrinsic transverse momentum (e.g. a Gaussian $\exp(-k_T^2/2\sigma^2)$) gets smeared out by the TMD evolution \cite{Martinez:2018jxt, Martinez:2019mwt}.  

\section{Conclusion} \label{sec:conclusion}
Two formalisms that include TMD physics, CSS and PB, have been systematically compared with respect to Sudakov resummation for DY $p_T$ spectra. At analytical level, CSS and PB coincide at LL and NLL order and differences at higher order coefficients are being tracked down. Numerically, both approaches are describing the low-$p_T$ region of the DY $p_T$ spectrum. Differences are observed in the estimation of theory uncertainties and the parametrization of nonperturbative effects. The features that emerged are potentially important for high-precision DY phenomenology and are well-suited within benchmark exercises for the LHC and future colliders.

\section*{Acknowledgements}
I am grateful to F.~Hautmann, T.~Cridge, F.~Coradeschi, M.~Pavlov, H.~Jung, A.~Lelek, S.~Catani, M.~Nefedov and L.~Keersmaekers for discussions and advice. 

\bibliographystyle{SciPost_bibstyle}

\bibliography{CSSvsPB}

\nolinenumbers

\end{document}